\documentclass[12pt]{article}
\pdfoutput=1

\usepackage{amsmath,amssymb,amscd}
\usepackage{listings}
\usepackage{caption}
\usepackage{dsfont}
\usepackage{slashed}
\usepackage{color}
\usepackage{ulem}

\usepackage[pdftex]{graphicx}
\usepackage{epstopdf}
\usepackage{subfigure}
\usepackage{epsfig}
\usepackage{listings}
\usepackage{caption}
\usepackage{cite}

\usepackage{multirow}

\newcommand{\beq}{\begin{equation}}
\newcommand{\eeq}{\end{equation}}
\newcommand{\nbea}{\begin{align*}}
\newcommand{\neea}{\end{align*}}
\newcommand{\nbeq}{\begin{equation*}}
\newcommand{\neeq}{\end{equation*}}

 \newcommand{\column}[1]{\left(\begin{array}{c} #1 \end{array}\right) }

 \usepackage{multirow}
\usepackage{array}
\newcolumntype{M}[1]{>{\centering\arraybackslash}m{#1}}
\newcolumntype{N}{@{}m{0pt}@{}}

\numberwithin{equation}{section}

\begin{document}


\pagestyle{empty}

\baselineskip=21pt
\rightline{{\fontsize{0.40cm}{5.5cm}\selectfont{KCL-PH-TH/2016-04, LCTS/2016-03, CERN-TH/2016-026}}}
\rightline{{\fontsize{0.40cm}{5.5cm}\selectfont{CAVENDISH-HEP-16-02, DAMTP-2016-20}}}
\vskip 0.75in

\begin{center}

{\large {\bf The Price of an Electroweak Monopole}}

\vskip 0.5in

 {\bf John~Ellis}$^{1,2}$,~
   {\bf Nick~E.~Mavromatos}$^{1,2}$
and {\bf Tevong~You}$^{3}$

\vskip 0.5in

{\small {\it

$^1${Theoretical Particle Physics and Cosmology Group, Physics Department, \\
King's College London, London WC2R 2LS, UK}\\
\vspace{0.25cm}
$^2${Theoretical Physics Department, CERN, CH-1211 Geneva 23, Switzerland}\\
\vspace{0.25cm}
$^5${Cavendish Laboratory, University of Cambridge, J.J. Thomson Avenue, \\ Cambridge, CB3 0HE, UK;\\
\vspace{-0.25cm}
DAMTP, University of Cambridge, Wilberforce Road, Cambridge, CB3 0WA, UK}
}}

\vskip 0.5in

{\bf Abstract}

\end{center}

\baselineskip=18pt \noindent


{\small
In a recent paper, Cho, Kim and Yoon (CKY) have proposed a version of the SU(2) $\times$ U(1) Standard Model
with finite-energy monopole and dyon solutions. The CKY model postulates that the effective U(1) gauge coupling $\to \infty$ 
very rapidly as the Englert-Brout-Higgs vacuum expectation value $\to 0$, but in a way that is incompatible with 
LHC measurements of the Higgs boson $H \to \gamma \gamma$
decay rate. We construct generalizations of the CKY model that are compatible with the $H \to \gamma \gamma$ constraint,
and calculate the corresponding values of the monopole and dyon masses. We find that the monopole mass could be
$< 5.5$~TeV, so that it could be pair-produced at the LHC and accessible to the MoEDAL experiment.
}


\vskip 0.75in

\leftline{ {February 2016}}

\newpage
\pagestyle{plain}

\section{Introduction}

Ever since Dirac first considered the possible existence of monopoles in QED~\cite{Dirac:1931kp}, and Schwinger
extended his considerations to dyons~\cite{Schwinger:1969ib}, theorists have explored the possible existence of
finite-energy monopoles and dyons, and tried to estimate their masses. As pointed out by
't Hooft~\cite{'tHooft:1974qc} and Polyakov~\cite{Polyakov:1974ek}, 
one very plausible scenario is that QED is embedded in a semi-simple unified group
with coupling $g_U$, in which case the core of the monopole/dyon is regularized and its mass is finite and ${\cal O}(V)/g_U$,
where $V$ is the vev of an Englert-Brout-Higgs field that breaks the unified group into pieces including a U(1)
factor with a U(1)$_{EM}$ component. 

However, physics at the electroweak scale is very well
described by the Standard Model, which has an SU(2) $\times$ U(1) group structure that does
not admit a finite-energy monopole or dyon solution unless its structure is modified~\cite{plb97,baecho}, and there is no sign of an underlying 
semi-simple unified group that might be broken down to the Standard Model at any accessible
energy scale. The question therefore arises whether there is any modification of the
Standard Model that might contain a monopole or dyon solution with a mass ${\cal O}(v)/g$,
where $g$ is a Standard Model gauge coupling and $v$ the vev of the Standard Model Englert-Brout-Higgs field.

Cho, Kim and Yoon (CKY)~\cite{cho2015} have recently proposed a scenario for modifying the
Standard Model that includes a non-minimal coupling of its Englert-Brout-Higgs field to the square of its U(1)
gauge coupling strength: (- 1/4) ${\cal L} \ni \epsilon(|H|/v) B_{\mu \nu} B^{\mu \nu}$. The coupling function
is normalized so that $\epsilon(|H|/v) \to 1$ as $|H| \to v$, in order to restore the conventional
normalization of the U(1) gauge field in the standard electroweak vacuum. Also, in order to have
a finite-energy dyon solution, the coupling function should vanish as $|H| \to 0$ like
$|H|^{n}: n > 4 + 2\sqrt{3} \simeq 7.46$, so as to regularize the energy integral at the origin.
Effectively, CKY create the possibility of a finite-energy dyon by postulating that the effective
U(1) gauge coupling $\to \infty$ sufficiently rapidly as $|H| \to 0$.

CKY do not discuss an ultraviolet completion of the Standard Model that might lead to such behaviour,
and nor do we. Our interest is limited to the question whether, in principle, the monopole mass could be regularized with
a value low enough for it to be pair-produced at the LHC, and hence accessible to the MoEDAL experiment~\cite{MoEDAL}.

In their original model, CKY postulated a simple power law for the coupling function: $\epsilon(|H|/v) \propto (|H|/v)^8$,
and calculated a dyon mass $M_D \simeq 0.65 \times (4\pi/e^2) M_W \simeq 7.2$~TeV.
There is, however, an experimental problem with this simple power-law Ansatz, since it leads to an effective
$H \gamma \gamma$ coupling that is much larger than is allowed by LHC measurements~\cite{ATLAS+CMS}. In the Standard
Model, the $H \gamma \gamma$ vertex is generated by loop diagrams (principally those involving
$W$ bosons and $t$ quarks), and hence is ${\cal O}(\alpha_{EM}/4 \pi)$. The data from CMS and ATLAS
on the $H \to \gamma \gamma$ decay rate~\cite{ATLAS+CMS} are quite consistent with this Standard Model calculation,
so they constrain any additional contribution to be ${\cal O}(10^{-3})$: see~\cite{esy}, for example. This implies that, if one
expands the coupling function $\epsilon(|H|/v)$ around the standard electroweak vacuum with
$|H| = v$, the linear term in the expansion, i.e.,
$\epsilon^\prime(|H|/v)|_{|H| = v}$, should be
${\cal O}(10^{-3})$~\footnote{We revisit this constraint more quantitatively in the following, but the
precise value is not very important for our estimate of the possible monopole mass.}.
This condition is manifestly not satisfied if $\epsilon(|H|/v)$ is a simple power of $|H|/v$,
but could be satisfied if $\epsilon(|H|/v)$ has a more complicated functional form.

We consider in this paper forms for $\epsilon(|H|/v)$ that contain various combinations of powers
$(|H|/v)^n: n \ge 8$, imposing the normalization condition $\epsilon(1) = 1$ and the LHC condition $\epsilon^\prime(1) = {\cal O}(10^{-3})$. 
If the form of $\epsilon(|H|/v)$ contains just two terms with different powers $n$, their coefficients can be
determined using these two conditions, and one can use the
classical equations of the Standard Model to calculate the energy (mass) of the lowest-lying
monopole configuration. However, if the form of $\epsilon(|H|/v)$ includes more terms,
the coefficients cannot be determined. Instead, we use as an additional constraint the Principle of Maximum Entropy (PME)~\cite{antolin},
namely that the quantity
\begin{equation}
\label{entropy}
S(\epsilon) = - \int_0^1 dx\, \epsilon(x)\, {\rm ln}\, \epsilon(x)
\end{equation}
should be maximized in the space of possible coefficients. Once $S(\epsilon)$ is maximized, one can again use the
classical equations of the Standard Model to calculate the energy (mass) of the lowest-lying
monopole configuration for the corresponding form of the coupling function $\epsilon(x)$.

We consider several possible functional forms for $\epsilon(|H|/v)$, and calculate the corresponding
values of the monopole mass ${\cal M}$. For a combination of $(|H|/v)^{10}$ and $(|H|/v)^{12}$
consistent with the LHC $H \to \gamma \gamma$ decay rate, we find
${\cal M} = 6.2$~TeV, increasing to 6.6~TeV for a combination of $(|H|/v)^{8}$ and $(|H|/v)^{10}$,
with no further reduction for the maximum-entropy combination of $(|H|/v)^{8}$, $(|H|/v)^{10}$ and 
$(|H|/v)^{12}$. On the other hand, forms of $\epsilon (|H|/v)$ combining higher powers $n = 14$ and 16
(with a logarithmic correction) yield lower monopole masses $\sim 5.7$ $(5.4)$~TeV. We conclude
that the CKY monopole could indeed weigh $< 5.5$~TeV, so that pair-production at the LHC is an open 
possibility, opening up interesting perspectives for the MoEDAL experiment~\cite{MoEDAL}.

\section{Review of the Cho-Maison Monopole Solution}

Before discussing the CKY construction~\cite{cho2015} of a finite-energy monopole solution in the electroweak 
theory, we first review the structure of the (infinite-energy)
Cho-Maison monopole solution. The Cho-Maison electroweak monopole~\cite{plb97}
is a numerical solution of the Weinberg-Salam theory~\footnote{An analytical existence theorem for 
such monopole solutions can be established by appropriately adopting 
arguments by Yang~\cite{yang}.}. However, it suffers from 
a divergence in the energy due to a singularity at the centre of the configuration, $r \to 0$, 
where $r$ is the radial coordinate. As such, it cannot be considered as physical in the absence of a
suitable ultraviolet completion.
CKY~\cite{cho2015} proposed a mechanism for rendering integrable the divergence at the monopole
core, yielding a finite-energy solution that would be physical.

The starting-point of Cho and Maison and CKY is the Lagrangian describing the bosonic sector of the Weinberg-Salam theory, 
\begin{align}
\mathcal{L} &= -|D_\mu H|^2 - \frac{\lambda}{2}\left(H^\dagger H - \frac{\mu^2}{\lambda}\right)^2 - \frac{1}{4}F_{\mu\nu}F^{\mu\nu} - \frac{1}{4}B_{\mu\nu}B^{\mu\nu}  \nonumber \\
&= -\frac{1}{2}(\partial_\mu \rho)^2 - \frac{\rho^2}{2}|D_\mu \xi|^2 - \frac{\lambda}{8}\left(\rho^2 - \rho_0^2\right)^2 \nonumber \\ 
&\quad\quad  - \frac{1}{4}F_{\mu\nu}F^{\mu\nu} - \frac{1}{4}B_{\mu\nu}B^{\mu\nu} \, ,
\label{LWS}
\end{align}
where the $SU(2)_L \times U(1)_Y$ gauge-covariant derivative is defined as 
\begin{equation*}
D_\mu \equiv \partial_\mu - i\frac{g}{2}\tau^a A^a_\mu - i \frac{g^\prime}{2}B_\mu 	\, ,
\end{equation*}
and $H$ is the Englert-Brout-Higgs doublet. In the second line of (\ref{LWS}) this is written as $H = \frac{1}{\sqrt{2}}\rho\xi$, where $\xi^\dagger \xi = 1$, and we define $\rho_0 = \sqrt{2}\mu^2/\lambda = \sqrt{2}v$. The $U(1)_Y$ coupling of $\xi$ is essential
for its interpretation as a $CP^1$ field with non-trivial second homotopy, making possible a topologically-stable monopole solution
of the equations of motion~\cite{plb97}. 

Choosing the following Ansatz for the fields in spherical coordinates $(t, r, \theta, \phi)$,
\begin{eqnarray}
\rho & = & \rho(r),  
~~~\xi=i\left(\begin{array}{cc} \sin (\theta/2)~e^{-i\varphi}\\
- \cos(\theta/2) \end{array} \right),   \nonumber \\
\vec A_{\mu} & = & \frac{1}{g} A(r)\partial_{\mu}t~\hat r
+\frac{1}{g}(f(r)-1)~\hat r \times \partial_{\mu} \hat r, \nonumber\\
B_{\mu} & = & \frac{1}{g'} B(r) \partial_{\mu}t 
-\frac{1}{g'}(1-\cos\theta) \partial_{\mu} \varphi.
\label{ans1}
\end{eqnarray}
one can find spherically-symmetric field configurations corresponding to electroweak monopoles
and dyons~\footnote{We emphasize that the
U(1)$_Y$ gauge symmetry is essential for permitting the spherically-symmetric Ansatz (\ref{ans1}), because spherical 
symmetry for the gauge field involves embedding the radial 
isotropy group SO(2) into the gauge group, which requires the 
Higgs field to be invariant under the U(1) subgroup of SU(2). 
This is possible with a Higgs triplet, but not with a Higgs 
doublet \cite{Forg}. In fact, in the absence of the U(1)$_Y$ 
degree of freedom, the above Ansatz describes the SU(2)
sphaleron, which is not spherically symmetric \cite{manton}.}. 
With this Ansatz, the equations of motion take the form
 \begin{align}
 \ddot{\rho} + \frac{2}{r}\dot{\rho} - \frac{f^2}{2r^2}\rho &= -\frac{1}{4}(A-B)^2\rho + \lambda\left(\frac{\rho^2}{2}-\frac{\mu^2}{\lambda}\right)\rho  \, , \nonumber \\
 \ddot{f} - \frac{f^2-1}{r^2}f &= \left(\frac{g^2}{4}\rho^2-A^2 \right)f  \, , \nonumber \\
 \ddot{A} + \frac{2}{r}\dot{A} - \frac{2f^2}{r^2}A &= \frac{g^2}{4}\rho^2(A-B)  \, , \nonumber \\
 \ddot{B} + \frac{2}{r}\dot{B} &= -\frac{{g^\prime}^2}{4}\rho^2(A - B) \, .
 \label{eq:eqm}
 \end{align}
After an appropriate unitary gauge transformation $U$ such that $\xi \to U\xi = \column{0 \\ 1}$,
one may obtain the physical gauge fields by rotating through the electroweak mixing angle $\theta_W$, 
\begin{align}
W_{\mu} &=\dfrac{i}{g}\frac{f(r)}{\sqrt2}e^{i\varphi}
(\partial_\mu \theta +i \sin\theta_W \partial_\mu \varphi), \nonumber\\
A_{\mu}^{\rm EM} &= e\left( \frac{1}{g^2}A(r)
+ \frac{1}{g'^2} B(r) \right) \partial_{\mu}t  
-\frac{1}{e}(1-\cos\theta_W) \partial_{\mu} \varphi,  \nonumber \\
Z_{\mu} &= \frac{e}{gg'}\big(A(r)-B(r)\big) \partial_{\mu}t,
\label{ans2}
\end{align}
where the electric charge $e = g \sin \theta_W = g^\prime \cos \theta_W$. The simplest non-trivial solution to the equations of motion with $A(r) = B(r) = f(r) = 0$ and $\rho=\rho_0 \equiv \sqrt{2}\mu/\sqrt{\lambda}$ describes a charge $4\pi/e$ point monopole with
\begin{equation*}
A_\mu^\text{EM} = -\frac{1}{e}(1 - \cos \theta)\partial_\mu\varphi \, .
\end{equation*}
More general electroweak dyon solutions may be obtained for non-zero $A, B$ and $f$. For example, with the boundary conditions
\begin{eqnarray}
&\rho(0)=0,~~f(0)=1,~~A(0)=0,~~B(0)=b_0, \nonumber\\
&\rho(\infty)=\rho_0,~f(\infty)=0,~A(\infty)=B(\infty)=A_0,
\label{bc0}
\end{eqnarray}
where $0 \leq A_0 \leq e\rho_0$ and $0 \leq b_0 \leq A_0$, we may integrate numerically the equations to obtain solutions representing the Cho-Maison dyon with electromagnetic charges
\begin{eqnarray}
&q_e=-\dfrac{8\pi}{e}\sin^2\theta_{\rm w} \int_0^\infty f^2 A dr 
=\dfrac{4\pi}{e} A_1, \nonumber\\
&q_m = \dfrac{4\pi}{e},
\label{eq:Charge}
\end{eqnarray}
where $A_1$ is a constant coefficient parametrising the $1/r$ asymptotic behaviour of $A$. 

However, the Cho-Maison electroweak monpole and dyon~\cite{plb97} suffer from 
a non-integrable singularity in the energy density at the centre of the configuration 
when $r \to 0$.
This can be seen by calculating the total energy $E$ of the dyon  configuration, which has the form~\cite{cho2015}:
\begin{gather}
E=E_0 +E_1,  \nonumber \\
E_0=4\pi\int_0^\infty \frac{dr}{2 r^2}
\bigg\{\frac{1}{g'^2}+ \frac1{g^2}(f^2-1)^2\bigg\}, \nonumber\\
E_1=4\pi \int_0^\infty dr \bigg\{\frac12 (r\dot\rho)^2
+\frac1{g^2} \left(\dot f^2 +\frac{1}{2}(r\dot A)^2 + f^2 A^2 \right) \nonumber\\
+\frac{1}{2g'^2}(r\dot B)^2 
+\frac{\lambda r^2}{8}\big(\rho^2-\rho_0^2 \big)^2 \nonumber\\
+\frac14 f^2\rho^2
+\frac{r^2}{8} (B-A)^2 \rho^2 \bigg\}.
\label{eq:energyint}
\end{gather}
We see that, with the boundary conditions given by (\ref{bc0}), $E_1$ is finite but the first term of $E_0$ is divergent at the origin. 

\section{Finite-Energy Monopoles and Dyons}
 
 The recent article by Cho, Kim and Yoon (CKY)~\cite{cho2015} proposed, as one possibility,
 regularising the Cho-Maison monpole by modifying the Weinberg-Salam theory
in such a way that the equations of motion have a finite-energy solution. 
The proposed modifications may be viewed as arising from unspecified dynamics that modify the form
of the dielectric `constant' in front of the $U(1)_Y$ hypercharge gauge kinetic term to become a non-trivial
functional of the Englert-Brout-Higgs doublet, $\epsilon(H^\dagger H)$, a construction that preserves gauge invariance. 
Specifically, CKY considered the following form of effective Lagrangian that 
has a non-canonical kinetic term for the $U(1)_Y$ gauge field 
\begin{equation}
{\cal L}_\text{eff} = -|D _\mu H|^2
-\frac{\lambda}{2} \Big(H^\dagger H -\frac{\mu^2}{\lambda}\Big)^2 
-\frac{1}{4} \vec F_{\mu \nu}^2  
-\frac{1}{4} \epsilon \left(\frac{|H|^2}{v^2} \right) B_{\mu \nu}^2,
\label{eq:Leff}
\end{equation}
where $\epsilon(|H|^2/v^2)$ is a positive dimensionless function 
of the Englert-Brout-Higgs doublet that approaches one asymptotically as $|H| \to v$. 
Clearly $\epsilon$ modifies the permittivity of the $U(1)_Y$ 
gauge field, but the effective action still retains the 
$SU(2)\times U(1)_Y$ gauge symmetry. Moreover, since 
$\epsilon \rightarrow 1$ asymptotically, the effective 
action reproduces the Standard Model when the Englert-Brout-Higgs field adopts its canonical vacuum expectation value: $|H| = v$. 
However, the factor $\epsilon (|H|^2/v^2)$ effectively changes the 
$U(1)_Y$ gauge coupling $g'$ to a ``running" coupling 
$\bar g'=g' /\sqrt{\epsilon}$ that depends on $|H|$. This is because, with 
the rescaling of $B_\mu \to B_\mu/g'$, $g'$ changes 
to $g' /\sqrt{\epsilon}$. By choosing $\epsilon$ so that $\bar g' \to \infty$
as $|H| \to 0$, i.e., requiring $\epsilon$ to vanish at the 
origin, one can regularise the Cho-Maison monopole. 

Such an {\it ad~hoc} modification of the Standard Model is phenomenologically motivated 
as a way to render finite the energy integral, leading to a finite mass for the electroweak monopole. 
We leave open the question of how such a modification may occur in a `top-down' approach,
and pursue the question how light such a CKY monopole might be~\footnote{We note that 
such effective theories with a field-dependent 
permittivity appear in non-linear electrodynamics models
and in higher-dimensional unified theories, and have been studied 
in cosmology in attempts to explain the late-time acceleration of the Universe~\cite{prd87,prl92,babi}.}.
.

With such a regularisation, the equations of motion in the spherically-symmetric ansatz are modified to
 \begin{align}
 \ddot{\rho} + \frac{2}{r}\dot{\rho} - \frac{f^2}{2r^2}\rho &= -\frac{1}{4}(A-B)^2\rho + \lambda\left(\frac{\rho^2}{2}-\frac{\mu^2}{\lambda}\right)\rho  \, , \nonumber \\
 &+ \frac{\epsilon^\prime}{{g^\prime}^2}\left( \frac{1}{r^4} - \dot{B}^2\right)\rho \, , \nonumber \\
 \ddot{f} - \frac{f^2-1}{r^2}f &= \left(\frac{g^2}{4}\rho^2-A^2 \right)f  \, , \nonumber \\
 \ddot{A} + \frac{2}{r}\dot{A} - \frac{2f^2}{r^2}A &= \frac{g^2}{4}\rho^2(A-B)  \, , \nonumber \\
 \ddot{B} + \left(\frac{2}{r} + 2\frac{\epsilon^\prime}{\epsilon}\rho\dot{\rho} \right)\dot{B} &= -\frac{{g^\prime}^2}{4\epsilon}\rho^2(A - B) \, ,
 \label{eq:eqm2}
 \end{align}
where we have defined $\epsilon^\prime \equiv d\epsilon / d\rho^2$. The original proposal~\cite{cho2015}
for regulating the infinite-energy divergence was to consider a functional form 
\begin{equation}
\epsilon \simeq \left(\frac{\rho}{\rho_0}\right)^n \, ,
\label{eq:originalepsilon}
\end{equation}
where one must require $n > 4 + 2\sqrt{3}$ in order for certain terms in the equations of motion to 
vanish fast enough as $r \to 0$ that the energy remains finite. With the boundary conditions (\ref{bc0})
the solution at the origin behaves as
\begin{align*}
\rho \simeq c_\rho r^{\delta_{-}} \quad , \quad f \simeq 1 + c_f r^2 \quad , \\
A \simeq c_A r \quad , \quad B \simeq b_0 + c_B r^{2\delta_{+}} \quad ,
\end{align*} 
where $\delta_{\pm} = \frac{1}{2}(\sqrt{3} \pm 1)$, and behaves asymptotically towards infinity as 
\begin{align*}
\rho \simeq \rho_0 + \rho_1 \frac{\exp(-\sqrt{2}\mu r}{r} \quad , \quad f \simeq f_1 \exp(-\omega r) 	\quad , \\
A \simeq A_0 + \frac{A_1}{r} \quad , \quad B \simeq A + B_1 \frac{\exp(-\nu r)}{r} \quad ,
\end{align*}
where $\omega = \sqrt{(g\rho_0)^2/4 - A_0^2}$ and $\nu = \frac{1}{2}\sqrt{g^2 + {g^\prime}^2}\rho_0$. 

\begin{figure}
\centering
\includegraphics[scale=0.385]{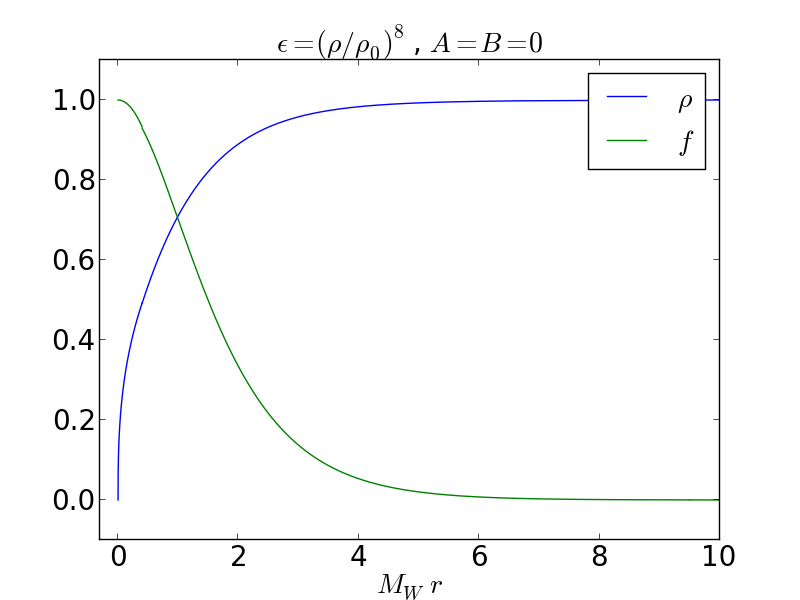}
\includegraphics[scale=0.385]{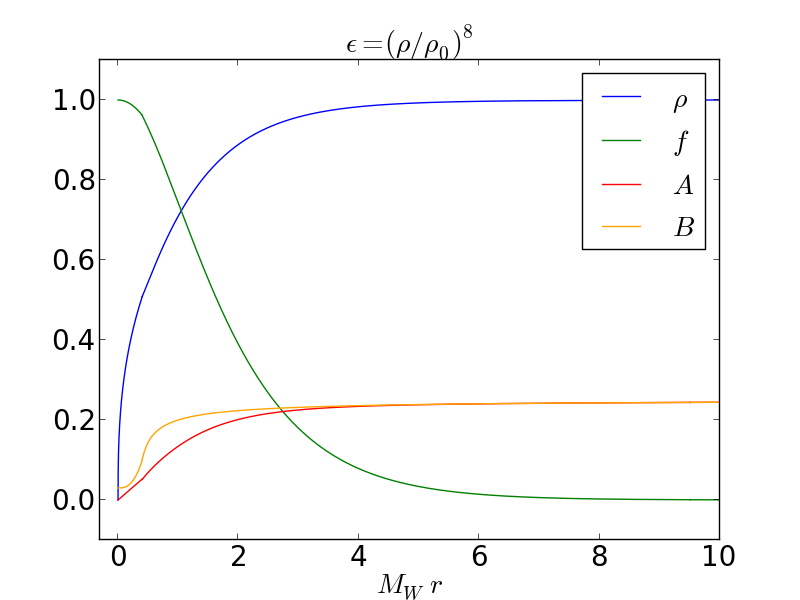}
\caption{Finite-energy electroweak monopole
solution for the $A=B=0$ case on the left and non-zero $A,B$ on the right with $\epsilon = (\rho/\rho_0)^n$: $n=8$. The $\rho$ and $f$ solutions are represented by solid blue and green lines, respectively, and the $A$ and $B$ fields
are denoted by red and orange lines, respectively. }
\label{fig:originalreg}
\end{figure}

These behaviours of the fields in the limits can be used together with the equations of motion (\ref{eq:eqm2}) to obtain numerical solutions.
We plot in Fig.~\ref{fig:originalreg} on the left the result for $n=8$ when $A=B=0$ (corresponding to a monopole with
no electric charge) and on the right the general case with $A_0 = M_W / 2$ (corresponding to a dyon). 
Plugging the simplest $A=B=0$ solution into the energy integral (\ref{eq:energyint}) with the appropriate $\epsilon$ form factor regularisation,  
we find a monopole mass of $\sim 5.7$ TeV. The non-zero $A, B$ solution yields a larger mass of $\sim 10.8$ TeV
for the dyon. An increase was to be expected, since non-vanishing forms of $A$ and $B$ will always contribute positively to the $E_1$ integral (\ref{eq:energyint}).

The topological stability of the lowest-lying monopole is guaranteed by the conservation of magnetic charge~\cite{plb97}.
However, dyon solutions may be unstable if suitable decays into charged particles and a monopole are
kinematically accessible, as is the case in this example.

\section{Phenomenological Constraint from $H \to \gamma \gamma$ Decay}

However, the simple power-law functional form for the $\epsilon$ regulator that was chosen
in~\cite{cho2015} is phenomenologically excluded by data on Higgs decays to $\gamma \gamma$~\cite{ATLAS+CMS}. 

In \cite{esy}, dimension-six operators involving couplings of the Higgs field with the gauge sector of the 
Standard Model have been studied in an analysis of the 
data now available from the LHC. Among them, of interest to us here is the operator
\begin{eqnarray}
\frac{c_\gamma}{\Lambda^2}\mathcal{O}_\gamma \equiv \frac{\bar{c}_\gamma}{M_W^2}{g^\prime}^2|H|^2B_{\mu\nu}B^{\mu\nu} \, ,
\label{eq:cgamma}
\end{eqnarray}
where we use the notation of Ref.~\cite{esy} in which constraints are placed on $\bar{c}_\gamma \equiv c_\gamma M_W^2/\Lambda^2$. Based on a global fit to LHC data, mainly from the decay of the Higgs field $H \to \gamma \, \gamma$, the best fit values of $\bar{c}_\gamma $ are in the range of $10^{-3}$ and negative~\cite{esy}.

Expanding $\rho $ near its vacuum expectation value  $\rho_0 \equiv \sqrt{2}\mu/\sqrt{\lambda}$:
\begin{equation}\label{rhoexp}
\rho = \rho_0 + \tilde \rho, \qquad \tilde \rho/\rho_0 \ll  1  \, ,
\end{equation}
we may write the term (\ref{eq:cgamma}) as an effective Lagrangian contribution of the form
\begin{align}
 \frac{\bar{c}_\gamma}{M_W^2}{g^\prime}^2|H|^2B_{\mu\nu}B^{\mu\nu} \supset 8\left(\frac{g^\prime}{g}\right)^2\bar{c}_\gamma\frac{\tilde\rho}{\rho_0}B_{\mu\nu}B^{\mu\nu} \, .
 \label{eq:effcgamma}
\end{align}
On the other hand, the $\epsilon$-dependent modification (\ref{eq:originalepsilon}) of the Lagrangian (\ref{eq:Leff}),
when expanded around the vacuum expectation value, yields a term
\begin{align}
-\frac{1}{4}\left(\frac{\rho}{\rho_0}\right)^2B_{\mu\nu}B^{\mu\nu} \supset -\frac{n}{4}\frac{\tilde\rho}{\rho_0}B_{\mu\nu}B^{\mu\nu} \, ,
\label{eq:effepsilon}
\end{align}
where we recall that finiteness of the monopole total energy/mass then requires for a simple
power law that $n \ge 8 \in Z^+$.

Comparing (\ref{eq:effcgamma}) with (\ref{eq:effepsilon}), we see that to linear order in $\tilde{\rho}/\rho_0$, 
\begin{align*}
\bar{c}_\gamma = -\frac{1}{32}\left(\frac{g}{g^\prime}\right)^2 n \simeq -0.1 n \, .
\end{align*}
Since $n \geq 8 \Rightarrow \bar{c}_\gamma \lesssim -0.8$ is strongly
excluded by the $95 \%$ CL observed value $\bar{c}_\gamma \gtrsim 10^{-3}$~\cite{esy},
we conclude that the simple power-law modification of the $U(1)_Y$ permeability proposed in~\cite{cho2015}
cannot be valid all the way from the origin of the Englert-Brout-Higgs field 
$\rho \to 0$  up to the region near the expectation value, $\rho \simeq \rho_0$. 

One needs therefore a modification of the Standard Model Lagrangian of the general form in (\ref{eq:Leff}), 
but with the $U(1)_Y$ permeability $\epsilon$ an interpolating functional having the following properties:
\begin{eqnarray} 
&& \epsilon(\rho) > 0 \nonumber \\
&&  \epsilon(\rho)|_{\rho =0} = \epsilon^{(1)} (\rho)|_{\rho =0}= \dots = \epsilon^{(n-1)}(\rho)|_{\rho =0} = 0 ~,   \nonumber \\
&& \epsilon^{(n)}(\rho)|_{\rho = 0} = \frac{n!}{\rho_0^n}  \ne 0~, \,  Z^+  \ni  n \ge 8~,\nonumber \\
&& \epsilon(\rho)| \simeq 1 - 16 \, \overline c_\gamma \Big(\frac{g^\prime}{g}\Big)^2 \frac{\rho^2}{\rho_0^2}
\; {\rm as} \; \rho \to  \rho_0~, \nonumber \\ && 
 |\bar{c}_\gamma| \lesssim {\mathcal O}(10^{-3}) \, ,
\label{eq:constraints}
\end{eqnarray} 
where the superscript $(n)$ indicates the $n$-th derivative with respect to $\rho$. In the following we impose the stronger condition 
$\bar{c}_\gamma = 0$: relaxing this to $|\bar{c}_\gamma| = {\cal O}(10^{-3})$ would not change our results significantly. 
\begin{figure}
\centering
\includegraphics[scale=1.2]{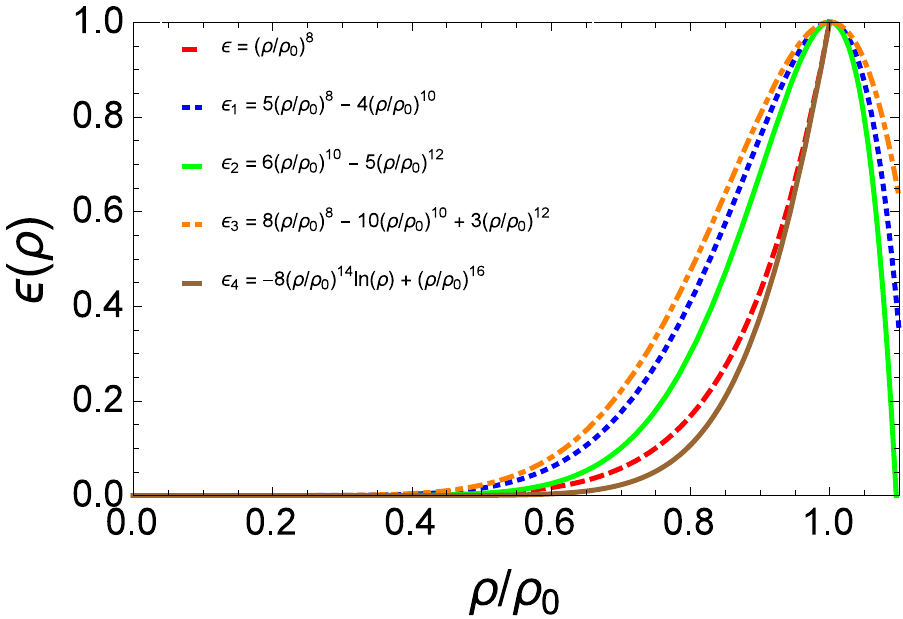}
\caption{Interpolating functions $\epsilon(\rho)$ that satisfy the required theoretical and phenomenological properties in solid brown, solid green, dotted blue, and dashed-dotted orange lines. The CKY regularisation~\protect\cite{cho2015}
that is incompatible with LHC data~\protect\cite{esy} is shown in dashed red. }
\label{fig:epsilon}
\end{figure}
\section{Implementing the $H \to \gamma \gamma$ Constraint}

\begin{figure}
\centering
\includegraphics[scale=0.385]{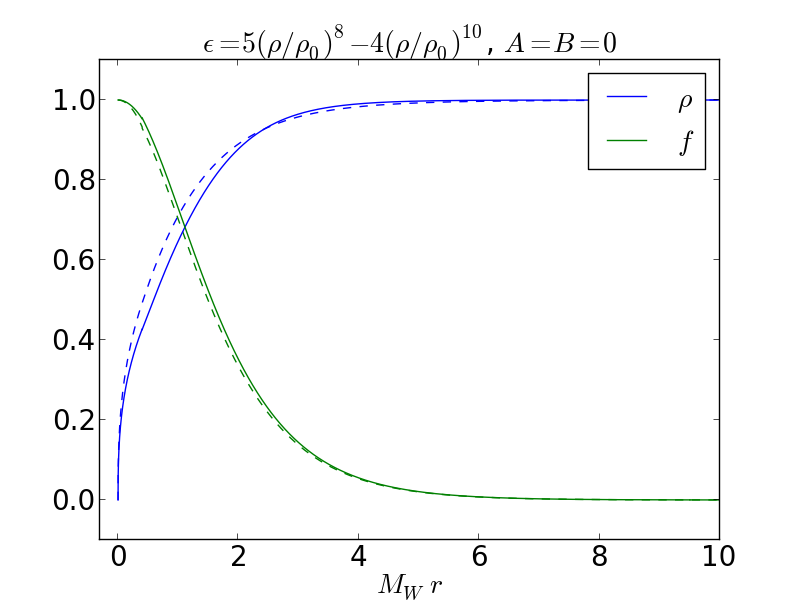}
\includegraphics[scale=0.385]{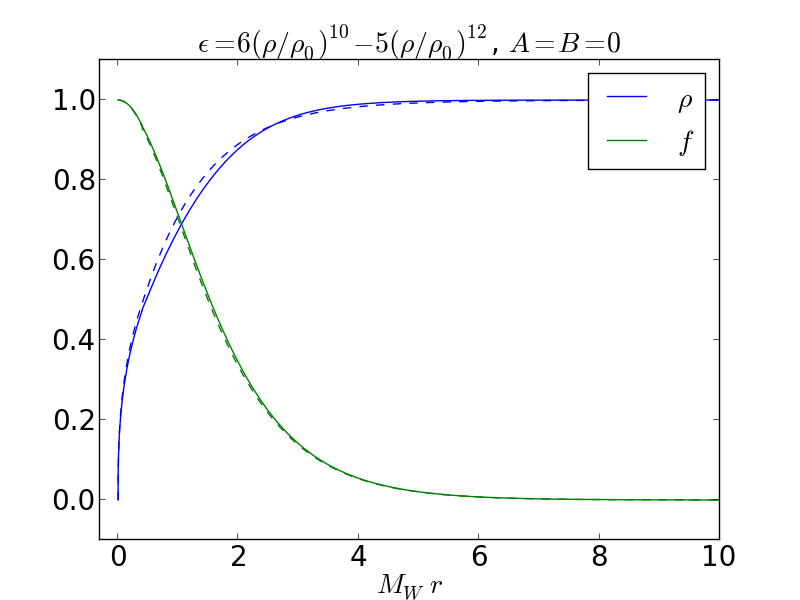}
\caption{Finite-energy electroweak monopole solutions obtained using the two-coefficient $\epsilon_1$ function
(\protect\ref{eps1}) on the left and $\epsilon_2$ function (\protect\ref{eps2}) on the right that satisfy all theoretical and phenomenological constraints. The solid blue (green) line 
represents the solution for $\rho$ ($f$), where $\rho$ is normalised by $\rho_0$. We also
plot the one-coefficient solution that is excluded by Higgs data using dashed lines. }
\label{fig:reg12ABzero}
\end{figure}

An acceptable form of the interpolating functional $\epsilon$ may be found by making an Ansatz with two
or more parameters, for which the simplest possibility is 
\begin{equation}
\epsilon_1(\rho) =  C_1 \left(\frac{\rho}{\rho_0}\right)^{8} + C_2 \left(\frac{\rho}{\rho_0}\right)^{10} \, .
\label{eps1}
\end{equation}
Solving for the coefficients $C_1$ and $C_2$ using the constraints (\ref{eq:constraints}) we find $C_1 = 5$ and $C_2 = -4$. 
This $\epsilon_1$ regularisation is plotted in dotted blue in Fig.~\ref{fig:epsilon}, with the original CKY $\epsilon$ 
denoted by a dashed red line for comparison. The solution for $\rho$ and $f$ with this $\epsilon_1$ regularisation is obtained 
numerically and plotted in the left panel of Fig.~\ref{fig:reg12ABzero}, 
and gives a monopole mass of $M \equiv E_0 + E_1 \simeq 6.6$ TeV 
when integrated in (\ref{eq:energyint}). We focus here on the $A=B=0$ monopole case,
as this minimises the total energy, with the lowest possible dyon mass being significantly larger.

The powers $n = 8, 10$ chosen in (\ref{eps1}) are the lowest powers of $\rho$ that are consistent with convergence
of the energy integral and analyticity in $|H|^2$. The monopole mass is larger than for the CKY Ansatz, because the
energy integrand must be larger at intermediate values of $\rho/\rho_0$ in order that $\epsilon$ be able to approach unity with
a very small derivative as $\rho \to \rho_0$. On the other hand, the fact that $n = 8$ is barely integrable suggests
that a smaller value of the monopole mass might be found for larger values of $n$.

Accordingly, we have examined a second Ansatz, $\epsilon_2$, that is a combination of $n = 10$ and $12$:
\begin{equation}
\epsilon_2(\rho) = 6\left(\frac{\rho}{\rho_0}\right)^{10} - 5 \left(\frac{\rho}{\rho_0}\right)^{12} \, ,
\label{eps2}
\end{equation}
where the values of the coefficients have again been chosen so that $\epsilon_2 \to 1$ with $\epsilon_2^\prime \to 0$
when $\rho \to \rho_0$.
The solution in this case is shown as a solid green line in Fig.~\ref{fig:epsilon}. 
Solving the equations of motion numerically once more,
we plot the result in the right panel of Fig.~\ref{fig:reg12ABzero}. Plugging this solution into the energy integral we find that the 
energy in this case is lowered to $M \simeq 6.2$ TeV, as anticipated because the regulating function
gives faster convergence as we approach the origin. 

\begin{figure}
\centering
\includegraphics[scale=0.385]{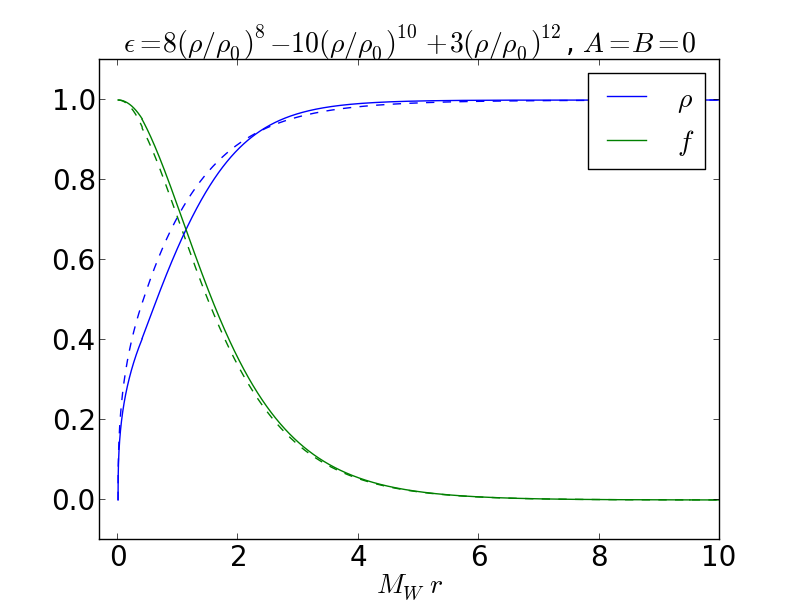}
\includegraphics[scale=0.385]{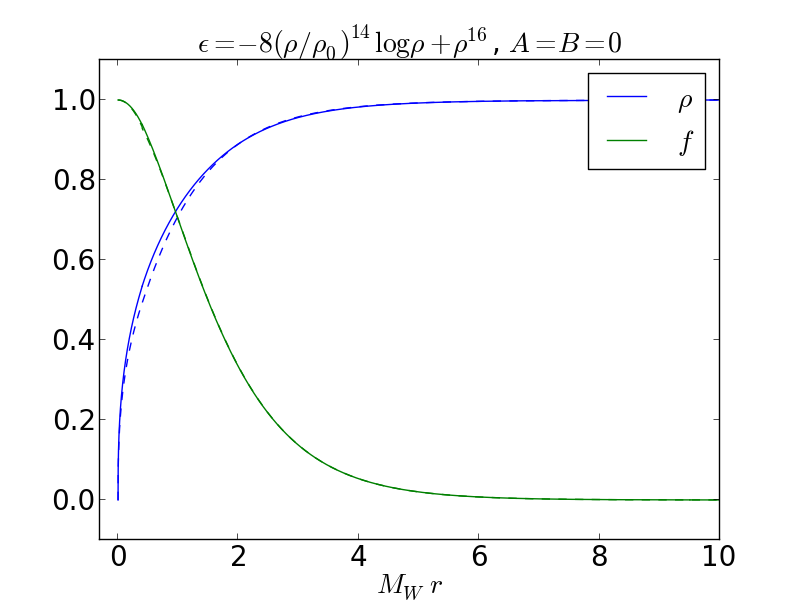}
\caption{Finite-energy electroweak monopole solution obtained using the three-coefficient $\epsilon_3$ function
(\protect\ref{eps3}) on the left and two-coefficient non-polynomial $\epsilon_4$ function (\protect\ref{eps4}) on the right that satisfy all theoretical and phenomenological constraints. The solid blue (green) line 
represents the solution for $\rho$ ($f$), where $\rho$ is normalised by $\rho_0$. We also
plot the one-coefficient solution that is excluded by Higgs data using dashed lines. }
\label{fig:reg34ABzero}
\end{figure}

One can also consider more complicated functional forms for $\epsilon$, that need not be polynomial
in $\rho$. Even if one considers just polynomials with more coefficients,
one has too many parameters to be determined by the number of conditions to be satisfied.
Finding the minimum of the energy integral over a multi-dimensional space is impractical.
Another possibility is to apply the principle of maximum entropy (PME) method~\cite{antolin}
to determine the coefficients. For example, one may consider the following Ansatz with three coefficients:
\begin{equation}
\epsilon_3(\rho) =  C_1 \left(\frac{\rho}{\rho_0}\right)^{8} + C_2 \left(\frac{\rho}{\rho_0}\right)^{10} + C_3 \left(\frac{\rho}{\rho_0}\right)^{12} \, ,
\end{equation}
where two combinations of coefficients can be determined by the two constraint equations obtained (\ref{eq:constraints}),
and one may solve for the remaining combination of coefficients by requiring that the entropy function 
\begin{equation}
S = -\int_0^1 dx \epsilon(x) \ln(\epsilon(x)) \, , \quad x \equiv \frac{\rho}{\rho_0} 
\end{equation}
be maximised. In this case we find the three-coefficient function that satisfies all these properties to be
\begin{equation}
\epsilon_3(\rho) =  8 \left(\frac{\rho}{\rho_0}\right)^{8} -10 \left(\frac{\rho}{\rho_0}\right)^{10} + 3 \left(\frac{\rho}{\rho_0}\right)^{12} \, ,
\label{eps3}
\end{equation}
which is plotted as an orange dash-dotted line in Fig.~\ref{fig:epsilon}. 
The numerical solution to the equations of motion for this three-coefficient $\epsilon_2$ 
regularisation is plotted for $A=B=0$ on the left in Fig.~\ref{fig:reg34ABzero}. 
Plugging this solution into the energy integral yields a slightly higher monopole mass than the two-coefficient
case, namely $M \simeq 6.8$ TeV. Thus, we do not find a lowering of the monopole mass with this simplest
generalisation to more coefficients of higher powers.

A lower monopole mass can be obtained by considering higher powers of $n$. For example, 
a two-coefficient polynomial regularisation with $n=14, 16$ yields a monopole mass of $M \simeq 5.7$ TeV.

One may also consider non-polynomial functional forms for $\epsilon$. As an example, we consider the following:
\begin{equation}
\epsilon_4(\rho) = -8\left(\frac{\rho}{\rho_0}\right)^{14} \log(\rho) + \left(\frac{\rho}{\rho_0}\right)^{16} \, ,
\label{eps4}
\end{equation}
which is plotted as a solid brown line in Fig.~\ref{fig:epsilon}. This regularisation converges faster,
due to the higher powers involved and the logarithm that modifies the behaviour of the function away from the 
vacuum expectation value of $\rho$ while vanishing when $\rho=\rho_0$. This solution is plotted in the right
panel of Fig.~\ref{fig:reg34ABzero}, and gives $M\simeq 5.4$ TeV. As expected from the improved convergence, this is the lowest monopole mass of all the $\epsilon$ regularisation functions that
we have considered here. The effect of the logarithm relative to the improvement due solely to the higher powers
$n=14, 16$ is seen in the reduction of the monopole mass from $M \simeq 5.7$ TeV. 

For the reader's convenience, our results for the specific $\epsilon$ regularisations that we have studied
are summarised in Table~\ref{tab:summary}. They lead to monopole masses ranging from $M \sim 6.8$ TeV
down to $\sim 5.4$ TeV, with a larger mass $\sim 10.8$~TeV for the dyon case we consider.
We may expect that a lower-mass monopole
could be found in a more exhaustive survey of parameter space,  particularly if attention was restricted to
powers $n \ge 10$.

\begin{table}
\centering
\begin{tabular}{| M{7cm} | M{2cm} |N }
\hline
$\epsilon$ regularisation & $M$ [TeV] & \\[5pt]  
\hline
\hline
$\left(\frac{\rho}{\rho_0}\right)^8$ & 5.7 &  \\[20pt] 
\hline
$\left(\frac{\rho}{\rho_0}\right)^8$ $(A, B \neq 0)$ & 10.8 &  \\[20pt] 
\hline
\hline
$ 5 \left(\frac{\rho}{\rho_0}\right)^{8} -4 \left(\frac{\rho}{\rho_0}\right)^{10}$ & 6.6 &  \\[20pt]  
\hline
$ 6 \left(\frac{\rho}{\rho_0}\right)^{10} - 5 \left(\frac{\rho}{\rho_0}\right)^{12}$ & 6.2 &  \\ [20pt] 
\hline
$  8 \left(\frac{\rho}{\rho_0}\right)^{8} -10 \left(\frac{\rho}{\rho_0}\right)^{10} + 3 \left(\frac{\rho}{\rho_0}\right)^{12}$ & 6.8 &  \\[20pt] 
\hline
$8\left(\frac{\rho}{\rho_0}\right)^{14} - 7 \left(\frac{\rho}{\rho_0}\right)^{16}$ & 5.7 &  \\[20pt]
\hline
$-8\left(\frac{\rho}{\rho_0}\right)^{14} \log(\rho) + \left(\frac{\rho}{\rho_0}\right)^{16}$ & 5.4 &  \\[20pt] 
\hline
\end{tabular}
\vspace{0.3cm}
\caption{ Monopole masses in TeV for the various $\epsilon$ regularisations that we consider. The first and second $\epsilon$ solutions are excluded by Higgs data while the rest satisfy all theoretical and phenomenological constraints listed in (\protect\ref{eq:constraints}).  } 
\label{tab:summary}
\end{table}

\section{Conclusion} 

Earlier papers by Cho and Maison~\cite{plb97} and by Cho, Kim and Yoon~\cite{cho2015} have indicated how
finite-mass electroweak monopole and dyon solutions may be found in suitable modifications
of the Standard Model. In particular, it was shown in~\cite{cho2015} that an appropriate
non-trivial permittivity in the U(1) sector of the Standard Model could regularise the
monopole and dyon energy integrals. However, the simplest example of such a scenario proposed
in~\cite{cho2015} is incompatible with data on $H \to \gamma \gamma$ decay from the LHC~\cite{ATLAS+CMS, esy}.
Nevertheless, we have shown in this paper how to generalise their construction in a way
that is compatible with these data and yields finite-mass monopole solutions.

The lowest monopole mass found in illustrative examples is $\simeq 5.4$~TeV,
and one may expect that a smaller mass could be found in a more complete study of generalisations of our
construction. However, a more exhaustive study should perhaps be contingent upon stronger
theoretical indications what type of modification of the U(1) permittivity might arise in which
completion of the Standard Model.

From our point of view, the most important conclusion of this exploratory study has been that
there is a possibility, compatible with the present constraints on the Standard Model~\cite{ATLAS+CMS,esy},
that there may exist an electroweak magnetic monopole with a mass
$< 5.5$~TeV, which could therefore be pair-produced at the LHC. We recall that the MoEDAL experiment~\cite{MoEDAL}
dedicated to searches for monopoles and other heavily-ionising particles has been installed at the LHC,
and has started taking data at 13~TeV in the centre of mass. Our analysis reinforces the
motivation to pursue the monopole search with MoEDAL.

\section*{Acknowledgements}

We thank Nick Manton and Jim Pinfold for their interest and relevant discussions.
The work of JE and NEM was supported partly by the London Centre for Terauniverse Studies (LCTS), using funding from the European Research Council via the Advanced Investigator Grant 26732, and partly by the STFC Grant ST/L000326/1. 
The work of TY was supported by a Junior Research Fellowship from Gonville and Caius College, Cambridge.

\end{document}